\documentclass[preprint,11pt]{aastex} 
\usepackage[]{epsfig}
\newcommand       \Angstrom     {\,{\rm \AA}}          
\newcommand	  \C		{\,{\rm C}}
\newcommand       \cm           {\,{\rm cm}}

\newcommand	  \g		{\,{\rm g}}
\newcommand	  \rmH		{{\rm H}}

\newcommand       \K            {\,{\rm K}}

\newcommand       \nH           {n_{\rm H}}
\newcommand       \bcnh         {b_{\rm C} N_{\rm H}}
\newcommand	  \pc		{\,{\rm pc}}

\newcommand       \simlt        {\lesssim}
\newcommand       \simgt        {\gtrsim}
\newcommand       \gtsim        {\gtrsim}
\newcommand       \ltsim        {\lesssim}
\newcommand       \fion         {\phi_{\rm ion}}        
\newcommand	  \um	        {\mu{\rm m}}
\newcommand	  \numC		{N_{\rm C}}	
\newcommand	  \bc		{{b_{\rm C}}}

\newcommand       \smcb         {{\rm SMC\ B1}\#1}

\newcommand	  \Urad		{U}
\newcommand	  \Umin		{U_{\rm min}}
\newcommand	  \Umax		{U_{\rm max}}
\newcommand	  \dN	        {dN_{\rm H}}

\newcommand       \magni        {\,{\rm mag}}
\newcommand       \ppm          {\,{\rm ppm}}
\newcommand	  \dof		{{\rm d.o.f.}}
\newcommand{\figwidth}{3.6in}

\countdef\decade=200
\advance\decade by \year
\countdef\hours=201
\advance\hours by \time
\divide\hours by 60
\countdef\mins=202
\advance\mins by \hours
\multiply\mins by 60
\multiply\hours by 100
\countdef\miltime=203
\advance\miltime by \hours
\advance\miltime by \time
\advance\miltime by -\mins

\shorttitle{Dust Emission in SMC}

\begin{document}

\title{
    \vspace*{-2.0em}
    {\normalsize\rm submitted to {\it The Astrophysical Journal}}\\
    \vspace*{1.0em}
        Infrared Emission from Interstellar Dust.\\
        III. The Small Magellanic Cloud
	}

\author{Aigen Li and B.T. Draine}
\affil{Princeton University Observatory, Peyton Hall,
        Princeton, NJ 08544, USA;\\
        {\sf agli@astro.princeton.edu, draine@astro.princeton.edu}}

\begin{abstract}
The infrared (IR) emission from interstellar dust in the 
Small Magellanic Cloud (SMC) is modelled using a mixture of 
amorphous silicate and carbonaceous
grains, including a population of polycyclic aromatic hydrocarbon (PAH) 
molecules.
(1) It is shown that this dust model  is able to reproduce 
the spectral energy distribution from near-IR to far-IR
for the entire SMC Bar region, provided the PAH abundance in the
SMC Bar region is very low.
(2) The IR spectrum of the \smcb\ molecular cloud can also be 
reproduced by our dust model provided the PAH abundance is increased
relative to the overall SMC Bar.  The PAHs in \smcb\ incorporate
$\sim$3\% of the SMC C abundance, compared to $<0.4\%$ in the SMC Bar.
(3) The spectrum of \smcb\ is best reproduced if the PAH mixture has 
intrinsic IR band strengths which differ from the band strengths 
which best fit Milky Way PAH mixtures.
The variation in the PAH IR band strengths would imply different
PAH mixtures, presumably a consequence 
of differing metallicity or environmental conditions.
Other possibilities such as super-hydrogenation of PAHs
and softening of the starlight spectrum are also discussed.
\end{abstract}


\section{Introduction\label{sec:intro}}

Due to its low metallicity ($\sim$1/10 of that in 
the Milky Way [Dufour 1984]) and low dust-to-gas ratio 
(over 10 times lower than in the Milky Way [Bouchet et al.\ 1985]), 
the Small Magellanic Cloud (SMC) is often considered as 
a prototype for the interstellar matter in high-redshift
galaxies at early stages of chemical enrichment.
The analogy is strengthened by the similarities of 
its typical extinction curve  -- a nearly linear rise 
with $\lambda^{-1}$ ($\lambda$ is wavelength)
and the absence of the 2175\AA\ hump 
(Lequeux et al.\ 1982; 
Pr\'evot et al.\ 1984; 
Bouchet et al.\ 1985;
Thompson et al.\ 1988; 
Rodrigues et al.\ 1997; 
Gordon \& Clayton 1998) -- 
to wavelength-dependent extinction inferred for 
starburst galaxies 
(Gordon, Calzetti, \& Witt 1997) 
and damped Ly$\alpha$ systems (Pei, Fall, \& Bechtold 1991). 

Infrared (IR) emission from dust in the SMC has been seen 
by the {\it Infrared Astronomical Satellite} (IRAS) 
(Schwering \& Israel 1990; Sauvage, Thuan, \& Vigroux 1990),
the Diffuse Infrared Background Experiment (DIRBE) 
instrument on the {\it Cosmic Background Explorer} (COBE) satellite 
(Stanimirovic et al.\ 2000), and the {\it Infrared Space Observatory} 
(ISO) (Reach et al.\ 2000). 
In the present work we use these observations to test the grain
model proposed previously to account for the observed extinction curve
for dust in the general SMC Bar region (Weingartner \& Draine 2001a).

Of particular interest is the detection 
by Reach et al.\ (2000)
of the 6.2, 7.7, 8.6, 11.3, and 12.7$\mu$m emission bands
toward the quiescent molecular cloud \smcb.
These bands, 
generally attributed to polycyclic aromatic hydrocarbons (PAHs) 
(L\'eger \& Puget 1984; Allamandola, Tielens, \& Barker 1985), 
had not previously been observed in the SMC.

The molecular cloud
\smcb\ ($[\alpha,\delta]_{1950}=[0^{\rm h}43^{\rm m}42^{\rm s}4,
-73^{\rm o}35^{\prime}10^{\prime\prime}]$)
has a 11.3$\mu$m/7.7$\mu$m emission band ratio which differs 
significantly from that for Milky Way regions, such as
the $\rho$ Oph reflection nebula (which one might expect 
SMC B1\#1 to resemble [see Reach et al.\ 2000]); 
the observed band ratio also falls outside the range 
which appears to be ``allowed'' by the PAH model 
of Li \& Draine 2001 (see Figures 16,17 of Draine \& Li 2001).
The observed spectrum for \smcb\ therefore presents an important
test which may reveal shortcomings 
of the current PAH emission models. The unusual band ratio
might be indicative of the physical conditions in \smcb, or it might
point to systematic differences between the PAH mixtures in the Milky
Way and SMC.

In this work we extend the interstellar dust emission model
developed for the diffuse interstellar medium (ISM)
of the Milky Way (Li \& Draine 2001; hereafter LD01) 
to the SMC, with emphasis on the mid-IR spectrum of \smcb. 
We seek
(1) to test the PAH IR emission model 
(Draine \& Li 2001; LD01);
(2) to infer the SMC PAH properties such as size distributions, 
charging, hydrogenation, and intrinsic IR band strengths;
(3) to infer the physical conditions in the SMC and 
the \smcb\ cloud and the dust properties in these environments.
Our approach is outlined in \S\ref{sec:dustmod} and \S\ref{sec:parameters}.
We first model the spectral energy distribution
for the SMC as a whole in \S\ref{sec:smcsed}.
Model results for the \smcb\ molecular cloud 
are presented in \S\ref{sec:mwmod}--\S\ref{sec:ir_band_strengths}.
In \S\ref{sec:extinction} we discuss the 2175\AA\ extinction hump
predicted from the present PAH model.
Our principal conclusions are summarized in \S\ref{sec:sum}.

\section{Modelling the Dust IR Emission\label{sec:model}}
\subsection{Dust Model\label{sec:dustmod}}

Following LD01, we model the SMC Bar and 
the \smcb\ molecular cloud IR emission as due to a mixture of
amorphous silicate grains and carbonaceous grains.
We assume that the carbonaceous grain population extends 
from grains with graphitic properties at radii $a\gtsim 0.01\micron$,
down to particles with PAH-like properties at very small sizes.  

Dust grains in the SMC and \smcb\ are assumed to be heated by 
a radiation field equal to the Mathis, Mezger, \& Panagia (1983; 
hereafter MMP) estimate of the solar neighbourhood interstellar 
radiation field (hereafter ISRF),\footnote{%
	The SMC Bar starlight spectrum may be harder
	than the MMP ISRF since the SMC has a higher surface density 
	of O and B stars (Lequeux 1989). On the other hand, grains within
	the \smcb\ cloud are exposed to a softer radiation field 
	due to reddening by dust in the cloud (see \S\ref{sec:reddenedisrf}).
	However, in comparison with other uncertainties, 
	these should be considered as second-order effects. 
	We will see in \S\ref{sec:reddenedisrf} that 
	reddening of starlight has only a minor effect 
	on the PAH mid-IR emission band ratios.
	}
multiplied by a factor $\Urad$.
The IR emissivity per unit solid angle per H nucleon from
a mixture of silicate and carbonaceous grains (including both 
neutral and ionized PAHs) is
\begin{eqnarray}
\nonumber
j_{\lambda}(\Urad) &=& \int^{\infty}_{3.5\Angstrom}
da \frac{1}{\nH}\frac{dn_{\rm carb}}{da}
      \int^{\infty}_{0} dT\ B_{\lambda}(T)\\
\nonumber
&&\times \Big\{ \fion(\Urad,a) C_{\rm abs}^{\rm carb^{+}}(a,\lambda)
\left(\frac{dP^{\rm carb^{+}}}{dT}\right)
      +
[1-\fion(\Urad,a)] C_{\rm abs}^{\rm carb^{0}}(a,\lambda)
\left(\frac{dP^{\rm carb^{0}}}{dT}\right)\Big\}\\
&&+ \int_{3.5\Angstrom}^\infty da \frac{1}{\nH}\frac{dn_{\rm sil}}{da}
\int^{\infty}_{0} dT\ B_{\lambda}(T)C_{\rm abs}^{\rm sil}
\left(\frac{dP^{\rm sil}}{dT}\right)
\label{eq:j_lambda}
\end{eqnarray}
where $n_{\rm H}$ is the hydrogen number density; 
$dn_{\rm carb}/da$ and $dn_{\rm sil}/da$
are the size distribution functions for carbonaceous and
silicate grains, respectively; 
$C_{\rm abs}^{\rm carb^{0}}$, 
$C_{\rm abs}^{\rm carb^{+}}$, and $C_{\rm abs}^{\rm sil}$
are the absorption cross sections for neutral and ionized 
carbonaceous grains (including PAHs) and silicate grains, respectively
(see LD01);
$B_{\lambda}(T)$ is the Planck function;
$dP^{\rm carb^{0}}(\Urad,a,T)$, 
$dP^{\rm carb^{+}}(\Urad,a,T)$, and 
$dP^{\rm sil}(\Urad,a,T)$ are, respectively,
the probabilities that the vibrational temperature 
will be in $[T,T+dT]$ for neutral and charged PAHs and silicate grains
illuminated by starlight intensity $\Urad$
(%
$dP/dT$ approaches a $\delta$ function for large grains);
and $\fion(\Urad,a)$ is the probability of finding a PAH molecule 
of radius $a$ in a non-zero charge state.

For a given radiation intensity $\Urad$, we calculate 
the temperature distribution functions $dP/dT$
employing the ``thermal-discrete'' method which treats 
both heating (photon absorption) and cooling (photon emission) 
as discrete transitions, using a thermal approximation for 
the downward transition probabilities (Draine \& Li 2001).
We have calculated $P(\Urad,a,T)$ for neutral and ionized PAHs
and silicate grains for $0.1\le \Urad \le 10^5$ at intervals 
of 0.25 in ${\rm log}_{10}\Urad$.

Except for single optically-thin clouds,
the region observed will generally include material exposed to a range of
radiation intensities.
We follow Dale et al.\ (2001) and 
assume the radiation intensity $\Urad$ to have a power-law
distribution of the general form
\begin{eqnarray}\label{eq:dNdU}
\frac{\dN}{d\Urad} &=& 
\frac{(1-\beta)N_{\rm H}^{\rm tot}}
{\Urad_{\max}^{1-\beta}-\Urad_{\min}^{1-\beta}}
\Urad^{-\beta} ~, ~~~~ {\rm for~}\Urad_{\min}\leq U \leq \Urad_{\max}
~~,~~\beta\neq 1 ~;
\\
&=& \frac{N_{\rm H}^{\rm tot}}{\ln(\Urad_{\max}/\Urad_{\min})}\Urad^{-1} ~,
~~~~{\rm for~}\Urad_{\min}\leq U \leq \Urad_{\max}
~~,~~\beta= 1 ~;
\end{eqnarray}
where $\Urad_{\min}$ and $\Urad_{\max}$ are, respectively,
the lower and upper cutoff of the radiation intensity $\Urad$,
and $N_{\rm H}^{\rm tot}$ is the total hydrogen column density.
Lacking knowledge of the detailed geometry of the dust and
illuminating stars, we will assume the different illumination
levels to be randomly distributed along the line-of-sight. 
The emergent IR intensity is then
\begin{equation}
\label{eq:I_lambda}
I_{\lambda} = 
\frac{1-{\rm exp}\left[-N_{\rm H}^{\rm tot} \Sigma_{\rm abs}(\lambda)\right]}
{N_{\rm H}^{\rm tot} \Sigma_{\rm abs}(\lambda)} 
\int_{\Urad_{\min}}^{\Urad_{\max}} 
d\Urad \frac{\dN}{d\Urad}j_{\lambda}(\Urad) 
~,
\end{equation}
where $\Sigma_{\rm abs}(\lambda)$ is 
the total {\it absorption}\footnote{%
  If the radiation field in the emitting region is
  approximately isotropic, scattering does not alter 
  the intensity.}
cross section per H nucleon for the dust model.
The integrated IR power per area is 
$N_{\rm H}^{\rm tot} p_0 \langle \Urad \rangle$,
where $p_0$ is the absorbed power per H nucleon for 
dust exposed to starlight with $\Urad=1$,
and the mean starlight intensity $\langle \Urad \rangle$ is 
\begin{eqnarray}\label{eq:power}
\langle \Urad \rangle &=& 
     \left(\frac{1-\beta}{2-\beta}\right)
     \frac{\Umax^{2-\beta}-\Umin^{2-\beta}}
     {\Umax^{1-\beta}-\Umin^{1-\beta}} ~,
     ~~~~ {\rm for~}\beta\neq 1,2 ~;
\\
&=& \frac{\Umax-\Umin}{\ln(\Umax/\Umin)} ~,
     ~~~~ {\rm for~}\beta=1 ~;
\\
&=& \frac{(1-\beta) \ln(\Umax/\Umin)}{\Umax^{1-\beta}-\Umin^{1-\beta}} ~,
     ~~~~ {\rm for~}\beta=2 ~. 
\end{eqnarray}

\subsection{Model Parameters\label{sec:parameters}}

We have three sets of parameters to be specified or constrained:
\begin{enumerate}
\item The column density $N_\rmH$ of gas averaged over the beam.
\item The parameters ($b_{\rm C}$, $a_0$, and $\sigma$)
      determining the abundance and
      size distribution of the ``log-normal'' population which
      contributes most of the ultrasmall carbonaceous particles (PAHs).
\item Environmental properties including 
      the starlight intensity $\Urad$ (determined by
      $\Urad_{\rm min}$, $\Urad_{\rm max}$, and $\beta$), 
      the electron density $n_e$, 
      and the gas temperature $T_{\rm gas}$.
      The heating and cooling of dust grains are determined 
      by $\Urad$. 
      The PAH ionization fraction $\fion$ depends on
      $\Urad/n_e$, and $T_{\rm gas}$.
\end{enumerate}

For dust in the general SMC Bar
we will adopt the grain size distribution obtained 
by Weingartner \& Draine (2001a) by fitting the extinction curve
toward the star AvZ 398 in the SMC Bar (Gordon \& Clayton 1998).
We will see that the PAHs in the SMC Bar make a negligible contribution
to the IR emission.
The SMC Bar models therefore depend 
on just 4 adjustable parameters:
$\Urad_{\min}$, $\Urad_{\max}$, $\beta$, 
and $N_{\rm H}^{\rm tot}$.

As noted by Reach et al.\ (2000), 
SMC B1\#1 may be considered to be an analog of
the $\rho$ Oph molecular cloud.
The dust in the $\rho$ Oph cloud has $R_V\approx 4.2$, so we will
assume that the dust grains in SMC B1\#1 have the
size distribution estimated by Weingartner \& Draine (2001a) for
Milky Way dust with $R_V=4.0$ ($\bc=20\ppm$, Case A), 
except 
(1) with abundances relative to H reduced by a factor 
10, the SMC gas-to-dust ratio relative to 
the local Milky Way value (Bouchet et al.\ 1985),
and
(2) we allow ourselves the freedom to adjust the parameters
$b_{\rm C}$, $a_0$, and $\sigma$.

We adopt a single log-normal size distribution for 
the PAHs in the \smcb\ cloud, 
characterized by three parameters: $a_{0}$, $\sigma$, and $\bc$;
$a_{0}$ and $\sigma$ respectively determine 
the peak location and the width of the log-normal distribution,
and $\bc$ is the total amount of C atoms relative to 
H locked up in PAHs.
Small PAHs (with $\ltsim 100$ carbon atoms) 
are expected to be planar (see Appendix A in Draine \& Li 2001).
The term ``PAH radius'' used in this and related papers
refers to the radius $a$ of a spherical 
grain with the same carbon density as graphite ($2.24\g\cm^{-3}$) and
containing
the same number of carbon atoms $N_{\rm C}$: 
$a \equiv 1.288 N_{\rm C}^{1/3}\Angstrom$. 

We adopt $n_{\rm H}\approx 10^3\cm^{-3}$, and
$T_{\rm gas}=15\K$ for the \smcb\ molecular cloud as
estimated by Lequeux et al.\ (1994) based on observed CO 
line ratios. 
After allowance for C in the dust grains, the gas phase C abundance
is [C/H]$_{\rm gas} \approx 3\times10^{-5}$.  In diffuse and translucent
regions with $A_V\ltsim 3$, we expect C and
other gas-phase metals (Mg, Si, S, Fe) to be photoionized, contributing
a fractional ionization $\sim 4\times10^{-5}$ in SMC gas.
We consider two ionization models: low $n_e$ and high $n_e$.
For the low-$n_e$ model we take $n_e=0.1\cm^{-3}$ 
($n_e/\nH\approx1\times10^{-4}$).
For the high-$n_e$ model we take $n_e=1.0\cm^{-3}$ 
($n_e/\nH\approx1\times10^{-3}$).
For the SMC Bar, we take $n_e=0.1\cm^{-3}$.
But this is not critical since the amount of PAHs 
in the SMC Bar is rather small and the PAH mid-IR 
emission is negligibly low comparing to the far-IR emission
(see \S\ref{sec:smcsed}). 

For a given $n_e$,
our model for SMC B1\#1 is then
determined by
7
parameters:
$\Urad_{\min}$;
$\Urad_{\max}$;
$\beta$;
the dust column $\propto N_{\rm H}^{\rm tot}$;
the fraction of the dust absorption due to PAHs $\propto \bc$;
and the parameters
$a_{0}$, $\sigma$ characterizing the PAH size distribution.
We will assume $\sigma=0.4$ for the width of the PAH size 
distribution as determined for the diffuse ISM PAHs (LD01).
Thus we are left with $N_{\rm par}=6$ adjustable parameters:
$\Urad_{\min}$, $\Urad_{\max}$, $\beta$, 
$N_{\rm H}^{\rm tot}$, $\bc$, and $a_0$.

\section{IR Emission from the SMC Bar and Eastern Wing\label{sec:smcsed}}
\subsection{Observational Constraints\label{sec:smcobs}}

We first consider the emission from the SMC as a whole.
In Figure \ref{fig:smc} we plot the surface brightness measured by DIRBE
for a 6.25 deg$^2$ region 
containing the optical bar and the Eastern Wing 
(Stanimirovic et al.\ 2000).
The overall spectrum, unlike that of \smcb\ (see \S\ref{sec:smcb1sed}),
peaks at $\lambda \sim 100\mu$m with a local minimum
at $\lambda\sim 12 \mu$m.

Stanimirovic et al.\ (1999) have mapped the HI 21cm emission from the SMC.
The total HI mass, corrected for self-absorption, 
is $\approx 4.2\times10^8M_\sun$ (Stanimirovic et al.\ 2000).
The 6.25 deg$^2$ region considered here contains 
$\approx 2.9\times10^8 M_\sun$ of HI 
(S. Stanimirovic 2001, private communication), 
corresponding to a mean HI column density 
$\langle N_{\rm HI}\rangle_{\rm obs}\approx 5.5\times10^{21}\cm^{-2}$,
consistent with 
$\langle N_{\rm HI}\rangle \approx 5.0\times 10^{21}\cm^{-2}$ 
determined by Lyman $\alpha$
observations of
21 sightlines
(Bouchet et al.\ 1985).
Based on a FUSE ({\it Far Ultraviolet Spectroscopic Explorer})
survey of H$_2$ along 26 lines of sight,
Tumlinson et al.\ (2002) found the H$_2$/HI mass ratio 
$M({\rm H_2})/M({\rm HI})\simlt 0.5\%$. 
However, Israel (1997) suggest
a global ratio $M({\rm H_2})/M({\rm HI})\approx 20\%$.  
We take 
$\langle N_\rmH\rangle_{\rm obs}\pm \sigma_{\rm H}=
(5.5\pm0.5)\times10^{21}\cm^{-2}$
for the region considered here.

For a model with $N_{\rm par}$ adjustable parameters, 
the goodness of fit is measured by
\begin{equation}
\frac{\chi^2}{\dof} = 
\frac{
([N_\rmH]_{\rm mod}-\langle N_\rmH\rangle_{\rm obs})^2/\sigma_\rmH^2
+
\sum_{i=1}^6 \left([\lambda I_{\lambda}]_{\rm mod}
-
[\lambda I_{\lambda}]_{\rm obs}\right)^2/\sigma_i^2 
}
{N_{\rm par}+7}
\end{equation}
measuring the difference between model and observations for the
6.25 deg$^2$ field studied by Stanimirovic et al.\ (2000).
We use observed 
12, 25, 60, 100, 140, and 240$\micron$ DIRBE band surface brightnesses
$[I_\lambda]_{\rm obs}$
and uncertainties $\sigma_i$
from Stanimirovic et al.\ (2000).

\subsection{Model Results\label{sec:smcrst}}

Grains over such a wide area will be illuminated 
by a range of radiation fields.
We model the SMC DIRBE spectrum 
assuming a power law distribution for the illuminating radiation
(see eq.\ [\ref{eq:dNdU}]), using the HI column density  
$\langle N_\rmH\rangle_{\rm obs}$ as an additional constraint.
As discussed in \S\ref{sec:parameters},
we are left with $N_{\rm par}=4$ adjustable parameters:
$\Umin$, $\Umax$, $\beta$, and $N_{\rm H}^{\rm tot}$.

\begin{figure}[h]
\begin{center}
\epsfig{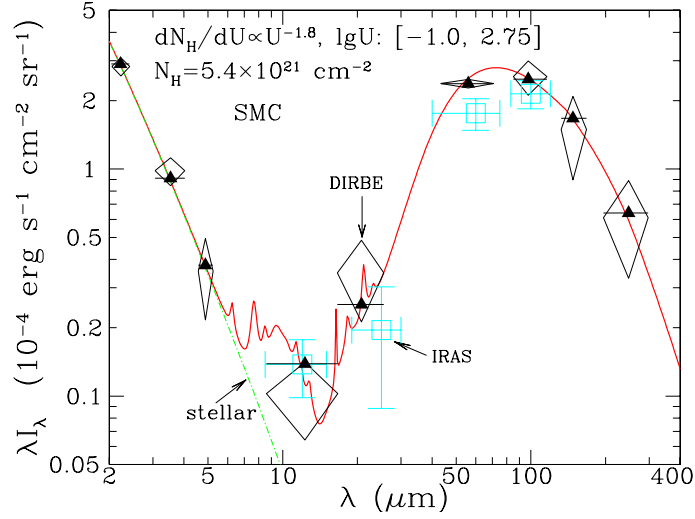}
\end{center}\vspace*{-1em}
\caption{
        \label{fig:smc}
        \footnotesize
	Comparison of the model (solid line) to the observed
        emission from the SMC obtained by COBE/DIRBE (diamonds)
        and IRAS (squares) averaged over a 6.25 deg$^2$ region
        including the optical bar and the Eastern Wing
	(Stanimirovic et al.\ 2000).
        Triangles show the model spectrum convolved with the DIRBE filters.
        Stellar radiation (dot-dashed line) dominates 
        for $\lambda\simlt 6 \mu$m. 
        Grains are illuminated by a range of
        radiation intensities $\dN/d\Urad\propto \Urad^{-1.8}$,
        $10^{-1.0} \le \Urad \le 10^{2.75}$,  
        with $N_{\rm H}^{\rm tot}\approx 5.4\times 10^{21}\cm^{-2}$.
        }
\end{figure}

Our best-fit model (see Table \ref{tab:para})
provides a good fit to the entire spectrum with 
emission from a mixture of silicate and carbonaceous grains with 
$N_{\rm H}^{\rm tot}\approx 5.4\times 10^{21}\cm^{-2}$,
illuminated by starlight with 
$\Umin=10^{-1.0}$, $\Umax=10^{2.75}$, $\beta=1.8$
(see Figure \ref{fig:smc}a).
The small value of $\chi^2/\dof=0.2$ suggests that some of the observational
uncertainties may have been overestimated.

\begin{table}[h,t]
{\tiny
\caption[]{Models for SMC Bar and SMC B1\#1.\label{tab:para}}
\begin{tabular}{lllllllllllllll}
\hline \hline
	Region
        & note
	& $n_e$\tablenotemark{a}
        & PAH band 

	& $a_{0}$ 
      	& $\sigma$ 
      	& $\bc$

      	& $\Urad_{\min}$
	& $\Urad_{\max}$
	& $\beta$
        & $\langle \Urad \rangle$
	& $N_{\rm H}^{\rm tot}$
      	& $\chi^2/$ 
        & $A_V$\tablenotemark{c}
      	& $\Delta A_{2175\Angstrom}$
	\\
        &
      	& ($\cm^{-3}$)
        & strength\tablenotemark{b}
	&(\AA) 
	& 
	& (ppm)
 
      	& (MMP)
	& (MMP)
	&
        & (MMP)
	& ($10^{21}\cm^{-2}$)
	& d.o.f.
        & (mag)
	&(mag) \\ 
\hline
Bar	& {\bf best} & {\bf0.01}& {\bf MW} & -- & -- & {\bf0}
	& {\bf10}$^{-1.0}$ & {\bf10}$^{2.75}$ & {\bf1.8} & {\bf1.9}& {\bf5.4}
        & {\bf0.2} & {\bf0.33} & --\\
Bar 	& $\Umin\uparrow$ &0.01& '' & -- & -- & ''
	& $10^{-0.5}$ & $10^{2.75}$ & 2 & 2.4& 4.8
        & 0.9 & 0.30 & --\\
Bar 	& $\Umax\downarrow$ &0.01& '' & -- & -- & ''
	& $10^{-1.0}$ & $10^{2.0}$ & 1.6 & 2.3& 4.8
        & 1.0 & 0.30 & --\\
Bar 	& $\beta\downarrow$ &0.01& '' & -- & -- & ''
	& $10^{-1.0}$ & $10^{1.75}$ & 1.5 & 2.4& 4.9 
        & 1.4 & 0.30 & --\\
Bar 	& $\beta\uparrow$ &0.01& '' & -- & -- & ''
	& $10^{0.25}$ & $10^{3.0}$ & 2.5 & 5.1& 2.4
        & 5.6 & 0.15 & --\\
Bar 	& single $U$ &0.01& '' & -- & -- & ''
	& $10^{1.5}$  & $10^{1.5}$& -- & 32& 0.28 
        & 10.2 & 0.02 & --\\
B1\#1   & {\bf best }&{\bf1.0} &{\bf SMC} &{\bf2.8} & {\bf0.4} & {\bf1.5}
        & {\bf10}$^{-0.5}$ &{\bf10}$^{2.5}$ & {\bf2} & {\bf2.2} & {\bf42} 
        & {\bf1.7} & {\bf2.05} & {\bf0.5} \\
B1\#1   & &0.1 & MW & 3 & 0.4 & 1.5
        & $10^{-0.5}$ & $10^{2.5}$ & 2 & 2.2 &$42$ 
        & 5.3 & $2.05$ &0.5 \\
B1\#1   & &1.0 & MW & '' & ''
        & '' & '' & '' & '' & '' & '' 
        & 3.1 & '' \\
B1\#1   & &0.1 & Eq.(\ref{eq:high2c}) & '' & '' & ''
        & '' & '' & '' & '' & '' 
        & 3.3 & '' &'' \\
B1\#1   & &1.0 & Eq.(\ref{eq:high2c}) & '' & '' & ''
        & '' & '' & '' & '' & '' 
        & 1.9 & '' &'' \\
B1\#1   & &0.1 & SMC & '' & '' & ''
        & '' & '' & '' & '' & '' 
        & 2.4 & '' &'' \\
B1\#1   & &1.0 & SMC & '' & '' & ''
        & '' & '' & '' & '' & '' 
        & 1.8 & '' &'' \\
B1\#1   & &0.1 & SMC & 2.8 & 0.4 & ''
        & '' & '' & '' & '' & '' 
        & 2.1 & '' &'' \\
\hline
\end{tabular}
\tablenotetext{a}{Electron density $n_e$ used to calculate
                  the PAH ionization fraction $\fion$,
		assuming $T_{\rm gas}=15\K$.}
\tablenotetext{b}{``MW'' refers to the PAH band strengths adopted
                  for the Milky Way diffuse ISM (LD01)
                  for which the 6.2, 7.7, and 8.6$\micron$ bands have 
                  been enhanced (relative to laboratory values) 
                  by factors of 3, 2, and 2, respectively;
                  and the H/C ratio is taken as Eq.(\ref{eq:nmlh2c});
                  ``Eq.(\ref{eq:high2c})'' refers to the same band
                   strengths as those of LD01 but
                   with a higher H/C (see Eq.[\ref{eq:high2c}];   
                   ``SMC'' refers to the laboratory measured 
                   band strengths except the 6.2$\micron$ band
                   is enhanced by a factor 1.5 
                   (with Eq.[\ref{eq:nmlh2c}]).}
\tablenotetext{c}{We adopt $A_V/N_{\rm H}^{\rm tot} = 
                  4.9\times 10^{-23}, 
                  6.2\times 10^{-23}\magni\cm^2$ 
                  respectively for the \smcb\ molecular cloud 
                  and the SMC Bar (Bouchet et al.\ 1985; 
                  Martin, Maurice, \& Lequeux 1989).}
}
\end{table}

Our best model has a 
dust-weighted mean radiation intensity $\langle U\rangle\approx 1.9$.
The observed mean surface brightness in the 1.25$\micron$ DIRBE band
averaged over the 6.25 deg$^2$ field
corresponds to only 0.76 of the intensity 
in the Mathis, Mezger, \& Panagia (1983) radiation field, 
but we expect $\langle U\rangle$ to be larger since
the dust is preferentially located in regions 
of high stellar density.\footnote{
	An enhanced radiation field ($\sim 10-100$ times 
	the Galactic mean value) is indicated by 
	the observed rotational excitation of H$_2$
	in the SMC diffuse ISM (Tumlinson et al.\ 2002).
	This may be due to the fact that the massive stars 
	and H$_2$ which FUSE observed are located near regions 
	of higher-than-average starlight.
}
Other distributions of radiation intensity are also able
to provide a fairly good fit to the observational data,
but only if $\Umin \ltsim 1$ and $\Umax\gtsim50$
(see Table \ref{tab:para}).

We have also modelled the SMC data using a single $\Urad$,
i.e., assuming that the dust is heated by a uniform radiation 
field with intensity $\Urad$. The best-fitting model is given
by $\Urad=10^{1.5}$ 
and $N_{\rm H}^{\rm tot}\approx 2.8\times 10^{20}\cm^{-2}$.
Although this model provides a fairly good fit to 
the overall spectrum, 
the derived hydrogen column density $N_{\rm H}^{\rm tot}$ 
is smaller than measured by a factor of $\approx 20$,
giving $\chi^2/\dof=10.2$.
This model is rejected:
the single $\Urad$ model has overestimated
the mean radiation field and correspondingly underestimated 
$N_{\rm H}^{\rm tot}$.

\subsection{The PAH Abundance in the SMC Bar\label{sec:smcbarpah}}

The SMC Bar dust model described above consumes 
$[{\rm C/H}]_{\rm dust} \approx 15\ppm$, and
$[{\rm Si/H}]_{\rm dust} \approx 12\ppm$.
Our SMC Bar grain model has a very low abundance of PAHs: 
the $a<15$\AA\ carbonaceous grains comprise only
$[{\rm C/H}]_{\rm PAH} \approx 0.2\ppm$, consistent with nondetection
of the 2175\AA\ extinction hump
(Weingartner \& Draine 2001a).

\begin{figure}[h]
\begin{center}
\epsfig{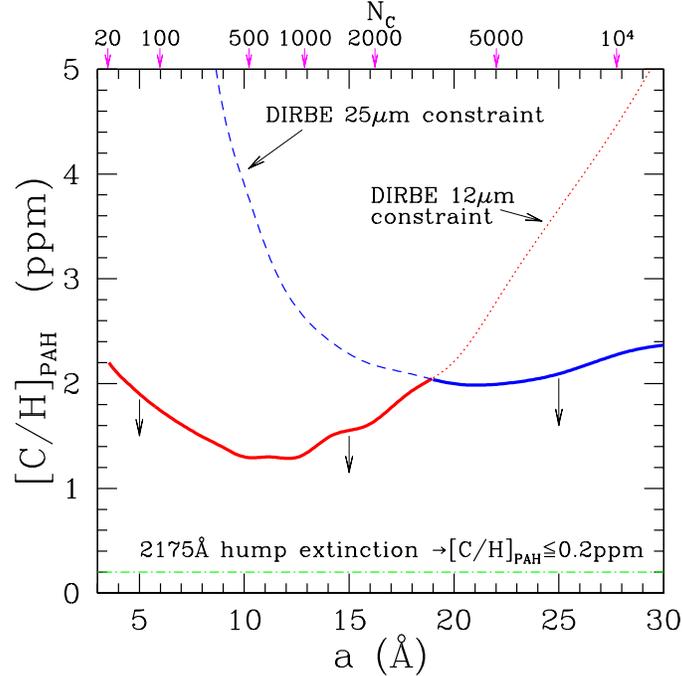}
\end{center}\vspace*{-1em}
\caption{
        \label{fig:upperlimits}
        \footnotesize
	Upper limits (solid line) on the abundances of PAHs in the SMC Bar 
	placed by 
        the DIRBE 12$\micron$ and 25$\micron$ 
        band photometry and the 2175\AA\ hump extinction (dot-dashed line).
        Overall, the DIRBE photometry limits the amount of 
        PAHs in the SMC Bar 
	to be $[{\rm C/H}]_{\rm PAH}\simlt 2\ppm$ (heavy solid line).
        The absence of the 2175\AA\ extinction hump provides 
        the strongest constraint: $[{\rm C/H}]_{\rm PAH}\simlt 0.2\ppm$
        (Weingartner \& Draine 2001a). 
        The upper axis indicates
        the number of C atoms $N_{\rm C}$ in a PAH molecule.
        }
\end{figure}

An upper limit on the PAH abundance can be obtained
by comparing the DIRBE 12$\micron$ and 25$\micron$ photometry
with theoretical IR spectra calculated for PAH molecules
illuminated by the above derived radiation field. 
We consider PAHs of a single size $a$.  
For each size $a$ we obtain an upper limit 
on $[{\rm C/H}]_{\rm PAH}$ in PAHs
such that this abundance of PAHs would not -- by itself --
exceed the DIRBE 12$\micron$ or 25$\micron$ photometry
(i.e., we ignore possible additional emission from silicate grains and
larger carbonaceous grains).
In Figure \ref{fig:upperlimits} we show the resulting upper limit 
$[{\rm C/H}]_{\rm PAH}$
as a function of PAH size $a$. 
It is seen that the DIRBE 25$\micron$ band places a stronger constraint 
than the 12$\micron$ band for PAHs $\simgt 19$\AA; 
whereas for smaller PAHs ($\simlt 19$\AA) the 12$\micron$ photometry
provides a stronger constraint.
Figure \ref{fig:upperlimits} also indicates that 
$a \approx 12$\AA\ grains are the best 12$\micron$ emitter:
the upper limits $[{\rm C/H}]_{\rm PAH}$ increase as grains
either increase or decrease their sizes.
In summary, we conclude that the DIRBE 12$\micron$ and 25$\micron$ 
photometry places an upper limit of
$[{\rm C/H}]_{\rm PAH}\simlt 2\ppm$ 
on the abundances of PAHs ($a<15$\AA).
This upper bound is much less stringent than the upper limit
$[{\rm C/H}]_{\rm PAH}\simlt 0.2\ppm$ 
implied by the absence of the 2175\AA\ extinction hump toward AvZ398.

The paucity of PAH molecules in the SMC Bar, 
together with the fact that the average extinction curve 
of the SMC has no observable 2175\AA\ hump,
is consistent with 
the hypothesis that PAHs are the carriers of the 2175\AA\ feature.
SIRTF
$5-25\micron$
observations of SMC regions where the 2175\AA\ feature is either
very weak or absent could strongly test our dust model, which predicts
minimal 7.7 and 11.3$\micron$ emission features from such regions (as
in Figure \ref{fig:smc}).

\section{IR Emission from the SMC B1\#1 Cloud\label{sec:smcb1sed}}

Having seen that our dust model is able to closely reproduce 
the emission from the SMC as a whole, we now consider the emission 
from the \smcb\ quiescent molecular cloud.

\subsection{Observational Constraints\label{sec:smcb1obs}}

The goodness of fit for \smcb\ is measured by
\begin{equation}
\frac{\chi^2}{\dof} = 
\frac{\sum_{i=1}^{156} W_i\left([\lambda I_{\lambda}]_{\rm mod}
-[\lambda I_{\lambda}]_{\rm obs}\right)^2/\sigma_i^2%
}
{N_{\rm par}+\sum_{i=1}^{156} W_i%
}
\end{equation}
where 
$N_{\rm par}$ is the number of adjustable parameters,
$[\lambda I_{\lambda}]_{\rm mod}$ is the model spectrum
(see Eqs.[\ref{eq:j_lambda},\ref{eq:I_lambda}]), and
$[\lambda I_{\lambda}]_{\rm obs}$ is the observed spectrum.

We use the 5--15$\micron$ ISOCAM spectrum (Reach et al.\ 2000).
For the 150 wavelengths in the ISO spectrum,
the weights
$W_i$ are taken to be
\begin{equation}
W_i = \left\{\begin{array}{ll} 
1,   &  ~~~{\rm for}\ |\lambda-\lambda_\alpha| < 
        \gamma_{\alpha}\lambda_\alpha\ 
        ~~(\alpha=1,2,...,5);\\
1/3, &  ~~~{\rm for}\ |\lambda-\lambda_\alpha| > 
        \gamma_{\alpha}\lambda_\alpha\
        ~~(\alpha=1,2,...,5);\\
1/6, & ~~~{\rm for}\ \lambda < 6.02\mu {\rm m}\ 
       ~{\rm or}~ \lambda > 13.38\mu {\rm m};\\
\end{array}\right.
\end{equation}
where $\lambda_\alpha$ and $\gamma_\alpha\lambda_\alpha$ 
are the peak wavelength and FWHM of the $\alpha$-th PAH feature
(see Table \ref{tab:pahparams}).
and $\sigma_i$ is the 1-$\sigma$ uncertainty 
in $[\lambda I_{\lambda}]_{\rm obs}$ at wavelength $\lambda_i$.

In addition, we use the IRAS 12, 25, 60, and 100$\micron$ intensities, 
with 
$\sigma_i=0.1(\lambda I_\lambda)_{\rm obs}$ (Schwering \& Israel 1990);
and the DIRBE 140 and 240$\micron$ photometry
(Hauser et al.\ 1998; Schlegel, Finkbeiner, \& Davis 1998)
after an empirical factor of 1.6 correction for beam dilution\footnote{%
	The DIRBE beam size ($0.7^{\rm o}\times 0.7^{\rm o}$) 
	is much larger than the angular size of the \smcb\ cloud 
	($\sim 50^{\prime\prime}\times 40^{\prime\prime}$,
	physical size $\sim 10\pc$), but there are a number of other
	emitting clouds in the DIRBE beam.
	To empirically correct for beam dilution, we multiply the 
	DIRBE 60, 100, 140, and 240$\micron$ surface brightnesses by a factor 
	of 1.6, bringing the DIRBE 60 and 100$\micron$ data into
	agreement with IRAS (see Figure \ref{fig:smcb1}). 
	This provides our best 
	estimate for the emission at 140 and 240$\micron$.%
	},
with $\sigma_i=0.2(\lambda/I_\lambda)_i$.
We assign $W_i=1$ for the 4 IRAS and 2 DIRBE bands, giving
$\sum W_{i=1}^{156} = 93$.

\subsection{Models with Milky-Way PAH Band Strengths\label{sec:mwmod}}
 
For our initial model, we use IR cross sections $C_{\rm abs}(a,\lambda)$ 
for neutral and charged PAHs following LD01.
We use the band strengths $\int S_\alpha(\lambda)d\lambda^{-1}$ 
recommended by LD01 for the Milky Way (MW) PAH mixture
where $S_\alpha(\lambda)$ is the absorption cross section per C atom, 
but we adjust central wavelengths $\lambda_\alpha$ and FWHMs 
$\gamma_\alpha\lambda_\alpha$
of the mid-IR features to match the observed spectrum.
Table \ref{tab:pahparams} shows the canonical MW
band positions and FWHM (from LD01) 
and the values appropriate for SMC B1\#1
(see also Table 1 in Reach et al.\ 2000).
Most of the changes are small, but the 11.3 and 12.7$\micron$ features
in \smcb\ are approximately twice as wide as for MW PAH mixtures,
while the $7.7\micron$ feature is $\sim$25\% narrower than in the MW. 

\begin{table}[h]
\caption[]{Drude Profile Parameters for PAHs
	\label{tab:pahparams}}
\begin{tabular}{ccccccc}
\tableline
&\multicolumn{2}{c}{$\lambda_\alpha(\mu{\rm m})$\tablenotemark{a}}&
\multicolumn{2}{c}{$\gamma_\alpha\lambda_\alpha(\mu{\rm m})$\tablenotemark{b}}&
\multicolumn{2}{c}{$E_\alpha$\tablenotemark{c}}\\
$\alpha$ &MW\tablenotemark{d}	
&SMC\tablenotemark{e}	
&MW\tablenotemark{d}	
&SMC\tablenotemark{e}	
&MW\tablenotemark{d}	
&SMC\tablenotemark{f}
\\
\hline
1	& 6.20	& 6.26		& 0.20	& 0.24 &3&1.5 \\
2	& 7.70	& 7.65		& 0.70	& 0.53 &2&1\\
3	& 8.60	& 8.48		& 0.40	& 0.36 &2&1\\
4	& 11.30	& 11.34		& 0.20	& 0.54 &1&1\\
5	& 12.70	& 12.80		& 0.30	& 0.58 &1&1\\
\tableline
\end{tabular}
\tablenotetext{a}{Central wavelength.}
\tablenotetext{b}{FWHM.}
\tablenotetext{c}{Enhancement of band strength
	$\int S_\alpha(\lambda) d\lambda^{-1}$ relative
	to average of lab values (see Table 1 of LD01).}
\tablenotetext{d}{Parameters recommended by LD01 for diffuse regions
	in the Milky Way.}
\tablenotetext{e}{Parameters giving good agreement with position and
	FWHM of 5-15$\micron$ features for
	the \smcb\ molecular cloud (see \S\ref{sec:mwmod}).}
\tablenotetext{f}{Band strengths giving good agreement with
	spectrum of the \smcb\ molecular cloud 
        (see \S\ref{sec:ir_band_strengths}).}
\end{table}

The PAH absorption cross sections $C_{\rm abs}(a,\lambda)$
and the temperature probability distribution functions $dP/dT$ 
depend upon the number of C atoms and the H/C ratio in the molecule.
As a starting point,
we take H/C to be
\begin{equation}\label{eq:nmlh2c}
{\rm H/C} = 
\left\{\begin{array}{lr} 
0.5, & \numC \le 25,\\
0.5/\sqrt{\numC/25}, & 25 \le \numC \le 100,\\
0.25, & \numC \ge 100,\\
\end{array}\right.
\end{equation}
which is appropriate for compact, symmetric PAHs
for $\numC \ltsim 10^2$, and clusters of such PAHs for
$\numC\gtsim 10^2$.
Draine \& Li (2001) found that the prescription (\ref{eq:nmlh2c})
appears to be consistent with the 3--15$\micron$ emission spectrum
seen in reflection nebulae and photodissociation regions in the Milky Way.

\begin{figure}[h]
\begin{center}
\epsfig{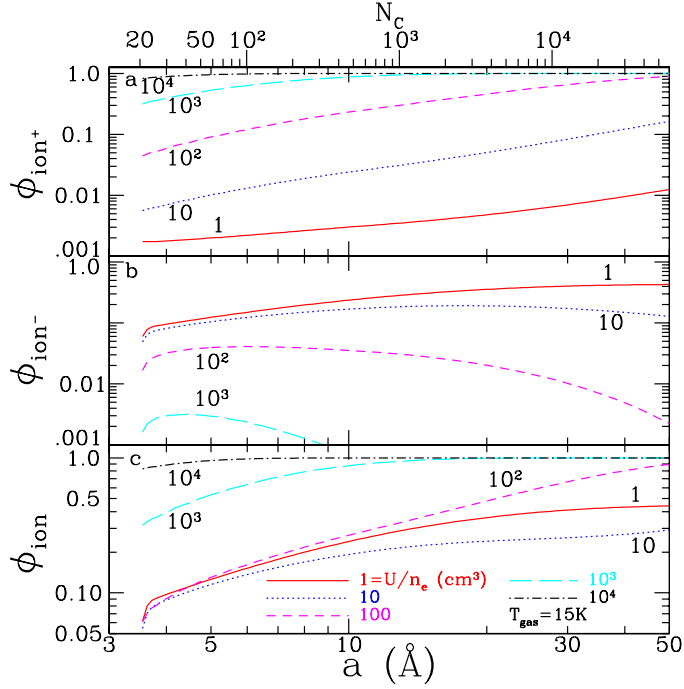}
\end{center}\vspace*{-1em}
\caption{
        \label{fig:ionfrac}
        \footnotesize
        The fraction of positively charged (a),
        negatively charged (b), and non-neutral (c) 
        PAHs as a function of size and radiation intensity
        for $n_e=0.1\cm^{-3}$ and $T_{\rm gas}=15\K$.
        The upper axis label gives the number of 
        carbon atoms $N_{\rm C}=0.468(a/\Angstrom)^3$. 
        }
\end{figure}

For a given $n_e$ and a given radiation intensity $\Urad$, 
we estimate the PAH ionization fractions 
$\fion (\Urad,a)$ in \smcb\ using the rates for 
photoelectric emission and electron capture recently 
discussed by Weingartner \& Draine (2001b).
Figure \ref{fig:ionfrac} shows
the ionization fractions calculated for 
a range of radiation intensities and PAH sizes.
At low $\Urad/n_e$, PAHs are more likely to be negatively charged 
(e.g., $\phi_{\rm ion^+}<\phi_{\rm ion^{-}}$ for $\Urad/n_e \simlt 10\cm^3$;
see Figure \ref{fig:ionfrac}a,b)
since the charging of a neutral PAH is dominated by electron capture.

\begin{figure}[h]
\begin{center}
\epsfig{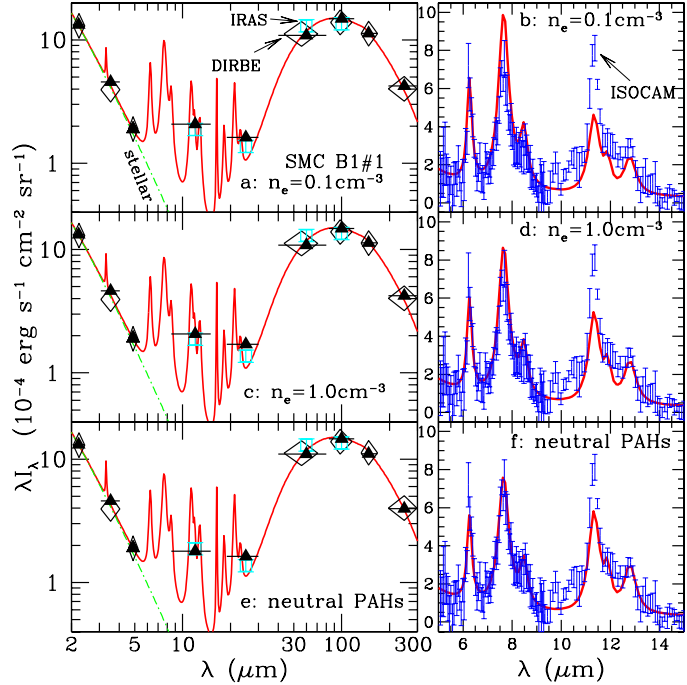}
\end{center}\vspace*{-1em}
\caption{
        \label{fig:smcb1}
        \footnotesize
	Comparison of the model (solid line) 
        to 
        IRAS photometry (squares), 
	DIRBE photometry (diamonds),
	and 5--15$\micron$ ISOCAM spectroscopy
        (vertical bars in panels b, d, and f)
        of the quiescent molecular cloud $\smcb$.
        Triangles show the model spectrum 
        convolved with the IRAS (12, 25, 60, 100$\micron$)
        and DIRBE (140, 240$\micron$) filters.
        Stellar radiation (dot-dashed line) is 
        approximated by a 6500$\K$ black-body.
        Upper panels (a,b): the best-fit low-$n_e$ dust model
        ($a_{0}=3$\AA, $\sigma=0.4$, $\bc =1.5\ppm$;
         $n_e=0.1\cm^{-3}$;
        $\chi^2/\dof\approx 5.3$) illuminated by a range of
        radiation intensities $\dN/d\Urad\propto \Urad^{-2}$,
        $10^{-0.5} \le \Urad \le 10^{2.5}$,  
        with 
        $N_{\rm H}^{\rm tot}\approx 
        4.2\times 10^{22}\cm^{-2}$.
        Middle panels (c,d): same as (a,b) but for
        the high-$n_e$ model ($n_e=1.0\cm^{-3}$;
        $\chi^2/\dof\approx 3.1$).
	Lower panels (e,f): same as (a,b) but for 
        a {\it pure neutral}-PAH dust model ($\chi^2/\dof\approx 1.9$).
        }
\end{figure}

For each of our two trial $n_e$ values,
we adjust $\Urad_{\min}$, $\Urad_{\max}$, $\beta$, $a_0$, 
$N_{\rm H}^{\rm tot}$, and $\bc$ to minimize $\chi^2$. 
In Figure \ref{fig:smcb1}a,b we plot the best-fitting
model spectrum for our low-$n_e$ model, for which
$\Umin=10^{-0.5}$, $\Umax=10^{2.5}$, $\beta=2$, 
$a_{0}=3$\AA, $\sigma=0.4$, 
$\bc = 1.5\ppm$, and
$N_{\rm H}^{\rm tot}\approx 4.2\times 10^{22}\cm^{-2}$
(parameters are summarized in Table \ref{tab:para}). 
Our estimate for the mean radiation intensity 
$\langle \Urad\rangle\approx 2.2$ is consistent with 
$\Urad \approx 3$ (with an uncertainty of a factor 2)
estimated by Reach et al.\ (2000).
We estimate that the dust in \smcb\ provides an extinction
$A_V\approx2.1$~mag.

Although the model closely reproduces 
the 12, 25, 60, 100, 140, and 240$\micron$ 
IRAS/DIRBE photometry (the 2.2, 3.5, and 4.9$\micron$ 
DIRBE bands are well represented by 6500$\K$ black-body 
stellar radiation), the model is not fully successful in 
reproducing the 6--13$\micron$ PAH emission features, resulting in the 
relatively large $\chi^2/\dof\approx 5.3$. 
In particular, it is a bit too weak for the 11.3$\mu$m 
C-H out-of-plane bending mode while too strong for the 
7.7$\mu$m C-C stretching mode.

In view of the fact that the 11.3$\mu$m feature is 
much stronger in neutral PAHs than in charged PAHs 
(Allamandola, Hudgins, \& Sandford 1999), 
the 11.3$\mu$m feature deficit could be alleviated 
by increasing $n_e$ so as to reduce 
the PAH ionization fraction (see Figure \ref{fig:ionfrac}).
Indeed, it is seen in Figure \ref{fig:smcb1}d 
that our high $n_e$ model provides an improved fit, 
with $\chi^2/\dof \approx 3.1$.
The unphysical case of {\it pure neutral} PAHs leads to a better fit
(see Figure \ref{fig:smcb1}f; $\chi^2/\dof \approx 1.9$),
but is still deficient in the 11.3$\micron$ band.
We discuss this further in \S\ref{sec:h2c}. 



For both low-$n_e$ and high-$n_e$ models, 
the number of C atoms (relative to H) in PAHs
required to account for the \smcb\ mid-IR spectrum
is $\bc \approx 1.5\ppm$.
The total C and Si abundances locked up in dust
($[{\rm C/H}]_{\rm dust} \approx 24.2\ppm$;
$[{\rm Si/H}]_{\rm dust} \approx 4.1\ppm$)
are consistent with 
the overall SMC elemental abundance
(C/H$\approx 54\ppm$, Si/H$\approx 11\ppm$,
Mg/H$\approx 9.6\ppm$,
Fe/H$\approx 6.9\ppm$;
Russell \& Dopita 1992).

Finally, we remark
that models with a single value of $\Urad$ are unable to
provide a satisfactory fit to the \smcb\ 60$\micron$ flux unless 
we add a log-normal PAH component peaking at 
$\sim 40-50$\AA~ containing a substantial mass
(e.g. ${\rm C/H} >8\ppm$ for $\Urad =3$).

\subsection{Models with Increased PAH Hydrogenation\label{sec:h2c}}

\begin{figure}[h]
\begin{center}
\epsfig{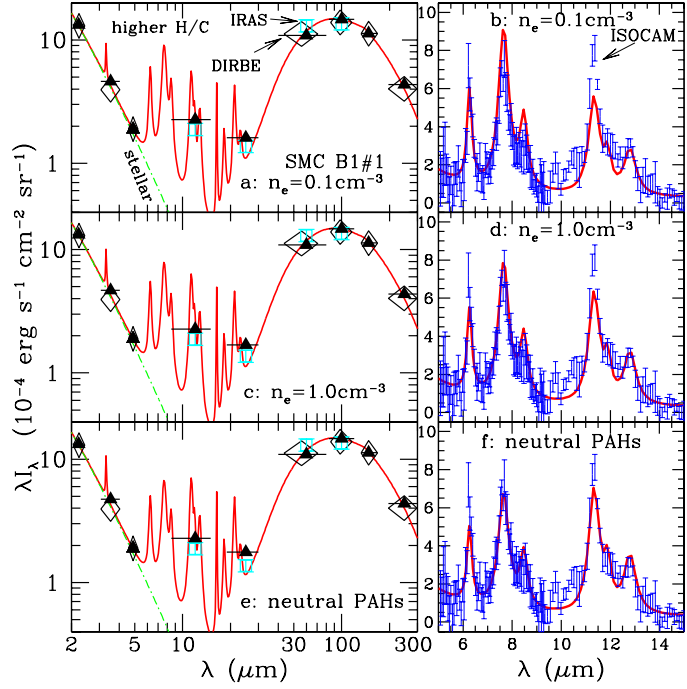}
\end{center}\vspace*{-1em}
\caption{
        \label{fig:hh2csmcb1}
        \footnotesize
	Same as Figure \ref{fig:smcb1}
        but for PAHs with a higher degree of hydrogenation
        (see Eq.[\ref{eq:high2c}]).
        }
\end{figure}

It is seen in Figure \ref{fig:smcb1} that even 
when entirely neutral our PAH model is still too weak
in the 11.3$\micron$ band.
Reach et al.\ (2000) suggested that the strong emission in 
the 11.3$\micron$ C-H bending mode might be due to increased 
hydrogenation of the SMC PAHs, which might perhaps be related to the 
fact that the gas phase C abundance (relative to H) is a factor 
of 10 lower than in the Milky Way (Dufour 1984; Russell \& 
Dopita 1992). To test this hypothesis, we have considered 
a PAH model 
where we assume the same intrinsic band strengths for C-C and C-H stretching
and bending modes, but 
with a higher H/C ratio (which would be appropriate 
for PAHs with more open structures):

\begin{equation}\label{eq:high2c}
{\rm H/C} = 
\left\{\begin{array}{lr} 
0.5, & \numC \le 100,\\
0.5/\sqrt{\numC/100}, & 100 \le \numC \le 400,\\
0.25, & \numC \ge 400.\\
\end{array}\right.
\end{equation}
We find that the best-fitting spectra 
(see Figure \ref{fig:hh2csmcb1}) do lead to 
improvements over those calculated from PAHs 
with a lower degree of hydrogenation (see Eq.[\ref{eq:nmlh2c}]).

However, this high H/C model still fails to attain
the observed 11.3$\micron$/7.7$\micron$ band ratio
even taking all PAHs to be neutral (see Figure \ref{fig:hh2csmcb1}c).
An increase in $a_0$ and/or $\sigma$ 
(implying a larger $\fion$ [see Figure \ref{fig:ionfrac}]
and a smaller H/C [see Eq.\ref{eq:high2c}]) does not help
since this would lead to a decrease at the 11.3$\micron$ band. 
While a slight decrease in $a_0$ and/or $\sigma$ 
(implying a smaller $\fion$ [see Figure \ref{fig:ionfrac}]
and a higher H/C [see Eq.\ref{eq:high2c}]) 
would result in an increase at the 11.3$\micron$ band,
this results in a decrease at the 6.2, and 7.7$\micron$ bands 
and thus does not work either since this model is already 
deficient in the 6.2 and 7.7$\micron$ bands 
(see Figure \ref{fig:hh2csmcb1}f).

In comparison with normal PAHs, super-hydrogenated PAHs would 
produce a stronger 3.3$\micron$ C-H stretching feature, 
a broad plateau below the 3.4 and 3.5$\micron$ features 
(Bernstein, Sandford, \& Allamandola 1996; Pauzat \& Ellinger 2001), 
and a stronger 6.2$\micron$ C-C stretching feature 
(Beegle, Wdowiak, \& Harrison 2001).
Future high-resolution observations of \smcb\, particularly at
the 3$\micron$ region, will provide us with more information
about the PAH hydrogenation. 

Since Milky Way PAHs with 
$\gtsim$25 C atoms are expected to be already 
essentially fully hydrogenated (Tielens et al.\ 1987; 
Allamandola, Tielens, \& Barker 1989; Allain, Leach, \& Sedlmayr 1996), 
an increase in the H/C ratio in the SMC would require that
the PAHs have more open structures than in the Milky Way.
There is no apparent
reason to expect this difference in PAH structure between 
the Milky Way and SMC.
Based on an analysis of the 10--15$\micron$ PAH out-of-plane 
C-H bending features, Vermeij et al.\ (2002) argue that 
the PAH emission in the \smcb\ cloud is dominated by
compact PAHs, i.e., relatively low hydrogenation.
In conclusion, it appears unlikely that 
the large 11.3$\micron$/7.7$\micron$ band ratio 
in \smcb\ is attributable to enhanced hydrogenation.


\subsection{Reddened Starlight?\label{sec:reddenedisrf}}

In previous sections the interstellar radiation field is taken
to have the same spectrum as the MMP solar neighbourhood field.
Although this assumption seems reasonable for the overall 
SMC where the extinction is small 
($A_V\approx 0.3\magni$; see \S\ref{sec:smcsed}), 
the starlight incident on the $\smcb$ molecular cloud surface 
will be subject to considerable reddening inside the cloud 
($A_V\approx 2\magni$; see \S\ref{sec:smcb1sed}).
The reddened starlight 
will affect the PAH emission bands in two ways: 
(1) PAH molecules will be excited to lower temperatures;
(2) the probability of being (positively) charged 
(via photoelectron ejection) will be smaller. 
One therefore would expect an increase in the
11.3$\micron$/7.7$\micron$ band ratio. 
Could reddening of the starlight 
explain the observed 11.3$\micron$/7.7$\micron$
band ratio in \smcb?

\begin{figure}[h]
\begin{center}
\epsfig{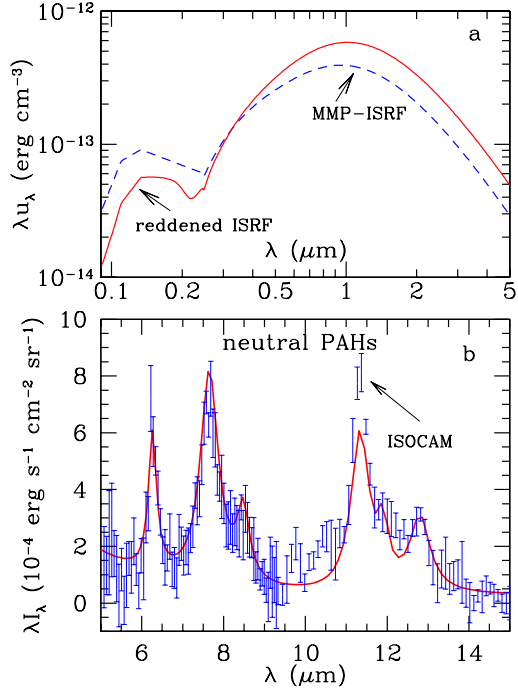}
\end{center}\vspace*{-1em}
\caption{
        \label{fig:reddenedisrf}
        \footnotesize
        Upper panel (a): trial reddened ISRF spectrum (solid line)
	with energy density equal to the MMP ISRF ($\Urad=1$).
        Also plotted is the MMP ISRF 
        ($\Urad=1$; dashed line).
        Lower panel (b): the best-fit pure neutral-PAH model
        ($a_0=2.6\Angstrom$, $\sigma=0.4$, $\bc=1.5\ppm$, 
        and $N_{\rm H}\approx 4.5\times 10^{22}\cm^{-2}$
        [for $\langle \Urad \rangle \approx 2.2$];
        $\chi^2/\dof\approx 1.9$).
        }
\end{figure}

To investigate this, we consider a trial reddened radiation field
with the spectrum shown in Figure \ref{fig:reddenedisrf}a,
with a factor of $\sim$3 reduction in the relative intensity
at $0.1\micron$ relative to $0.5\micron$.
We have calculated the temperature probability distribution functions
for PAHs excited by this radiation field with $\Urad=1$.
The mid-IR
emission spectrum is entirely due to single-photon heating, so the
emission per PAH simply scales as $\Urad$;
the spectrum in the lower panel of Figure \ref{fig:reddenedisrf} is
calculated for $\Urad=2.2$.
It is seen in Figure \ref{fig:reddenedisrf}b that
even for pure neutral PAHs, reddened starlight results in an emission
spectrum which is very similar to the pure neutral PAH spectrum in
the unreddened radiation field in Figure \ref{fig:smcb1}f, and is
unable to reproduce the observed 11.3$\micron$/7.7$\micron$ band ratio.
We conclude that excitation by reddened starlight 
cannot account for the 11.3$\micron$/7.7$\micron$ band ratio observed
toward \smcb.

\subsection{%
	IR Band Strengths: Milky-Way PAHs vs.\ SMC PAHs
	\label{sec:ir_band_strengths}}

Individual PAHs differ considerably in the relative strengths of
different vibration bands.
LD01 discussed laboratory measurements and theoretical
calculations of band strengths for selected PAH molecules and ions,
but argued that spectroscopy of various objects in the Milky Way
was best reproduced if the 6.2, 7.7 and 8.6$\micron$ 
band strengths were increased by factors of $E_{6.2}=3$, 
$E_{7.7}=2$, and $E_{8.6}=2$ respectively,
relative to the average laboratory and theoretical values.
These enhancement factors were used 
for the models shown in Figures \ref{fig:smcb1}--\ref{fig:reddenedisrf}.

It is reasonable to ask whether the observed spectrum indicates that
the band strengths for the PAHs in \smcb\ differ from the band strengths
which characterize PAHs in the Milky Way.  If so, what band strengths
give the best fit to the \smcb\ spectrum?

\begin{figure}[h]
\begin{center}
\epsfig{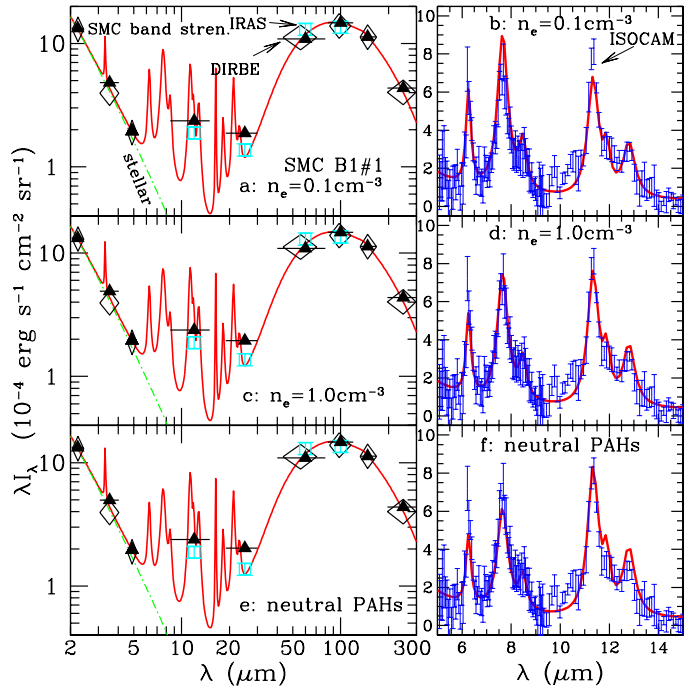}
\end{center}\vspace*{-1em}
\caption{
        \label{fig:e1smcb1}
        \footnotesize
	Same as Figure \ref{fig:smcb1}
        but for PAHs with ``SMC'' band strengths
	(see text)
        and a size distribution of $a_0$=2.8\AA\ and $\sigma$=0.4.
	We favor this model with $n_e\approx1.0\cm^{-3}$ for \smcb.
        }
\end{figure}

In Figure \ref{fig:e1smcb1} we show the emission calculated from a model with
$E_{6.2}=1.5$, $E_{7.7}=E_{8.6}=1$. For $n_e=1.0\cm^{-3}$ this model 
(see parameters in Table \ref{tab:para}) is in
very good agreement with the observations, with
$\chi^2/\dof=1.7$.
The agreement is considerably better than for the models considered in
Figures \ref{fig:smcb1}--\ref{fig:reddenedisrf}.
Note that these adjusted band strengths are in fact {\it closer} to
the average laboratory values than are the ``Milky Way'' band strengths.

Alternatively, 
the apparent need for a change in intrinsic band strength
might be due to the oversimplification
of the ``astronomical'' PAH model in which we do not
distinguish anions (PAH$^-$) and multiply-charged 
cations (PAH$^{n+}$) from singly-charged cations (PAH$^+$). 
Bauschlicher \& Bakes (2000) predict that
the 6.2, 7.7$\micron$ C-C stretching modes
and the 8.6$\micron$ in-plane C-H bending mode
are even more enhanced for PAH$^{n+}$ than PAH$^+$.
PAH$^-$ anions have band strengths intermediate between those of neutrals
(strong 3.3$\micron$ C-H stretching
and 11.3, 11.9, 12.7$\micron$ out-of-plane C-H bending modes)
and PAH$^+$ cations (strong 6.2, 7.7$\micron$ C-C stretching 
and 8.6$\micron$ C-H in-plane bending mode).
Unfortunately, no experimental measurements have yet been made
for multiply-ionized PAHs. 
Available evidence indicates that the
IR properties of PAH$^-$ anions closely resemble 
those of PAH$^+$ cations except for the very strong 
3.3$\mu$m C-H stretch enhancement in the anion 
(Szczepanski, Wehlburg, \& Vala 1995; Hudgins et al.\ 2000).


\subsection{UV Extinction\label{sec:extinction}}

Laboratory measurements show that PAHs have an absorption 
peak due to the $\pi$-$\pi^{\ast}$ electronic 
transition in the vicinity of 2000--2500\AA, and in the 
grain model considered here (LD01) it is
assumed that the interstellar PAH mixture contributes
absorption with the wavelength-dependence of the 2175\AA\
feature observed on sightlines in the Milky Way and the
Large Magellanic Cloud (LMC). The PAH abundance required 
to reproduce the observed IR emission features in the Milky Way
(C/H$\approx 60\ppm$) is such that we posit
that PAHs in fact contribute most of the observed 2175\AA\ feature.
The predicted extinction excess at 2175\AA\ is (see LD01)
\begin{equation}
\Delta A_{2175\Angstrom} \approx 0.85 
\left(\bcnh/10^{17}\C\cm^{-2}\right) {\rm mag} ~~~.
\end{equation}
Our best-fitting model has
$N_\rmH=4.2\times10^{22}\cm^{-2}$ and
$\bc =1.5\ppm$ 
(see \S\ref{sec:smcb1sed}).
Therefore we predict $\Delta A_{2175\Angstrom} \approx 0.5\magni$
for sightlines through \smcb.
We note that the $\Delta A_{2175\Angstrom}$ 
value derived here is a lower limit 
since $15\simlt a\simlt 150\Angstrom$ carbonaceous grains
may also contribute to the 2175\AA\ hump.

This prediction appears to conflict with the fact that most SMC 
extinction curves have no detectable 2175\AA\ hump 
(Pr\'evot et al.\ 1984). However, as 
pointed out by Reach et al.\ (2000), there may be
regional variations in the SMC extinction curve.
We note that the sight lines which show no 2175\AA\ hump 
all pass through the SMC Bar regions of active
star formation (Pr\'evot et al.\ 1984; Rodrigues et al.\ 1997; 
Gordon \& Clayton 1998). There is at least one line of sight 
(Sk 143$=$AvZ 456) with an extinction curve with a strong 
2175\AA\ hump (Lequeux et al.\ 1982; Pr\'evot et al.\ 1984; 
Bouchet et al.\ 1985; Thompson et al.\ 1988; Rodrigues et al.\ 1997; 
Gordon \& Clayton 1998). This sightline passes through the SMC wing,
a region with much weaker star formation (Gordon \& Clayton 1998).
It is possible that the \smcb\ dust, 
just like the dust toward Sk 143,
has been exposed to a less harsh environment than that in the 
SMC Bar regions where the UV extinction has been measured, and 
where the radiation and shocks associated with
massive stars might have destroyed PAHs.
Finally, we also note that {\it both} the 2175\AA\ hump 
(Fitzpatrick 1985) {\it and} PAH emission features 
(Sturm et al.\ 2000) are present in the 30 Dor region 
in the LMC, 
another low-metallicity environment 
($\sim$1/4 of that in the Milky Way; Dufour 1984).

\section{Conclusions\label{sec:sum}}

In contrast to the Milky Way Galaxy, the SMC is characterized 
by a lower metallicity, a lower dust-to-gas ratio, 
and an absence of the 2175\AA\ hump 
on most sightlines with measured UV extinction (see \S\ref{sec:intro}).
Therefore, the SMC provides an ideal laboratory 
to study dust in an environment quite different from the Milky Way.
We have modelled the IR spectral energy distribution 
for the SMC as a whole (averaged over a 6.25 deg$^2$ area)
(\S\ref{sec:smcsed}). 
We have also modelled the IR emission spectrum 
of the quiescent molecular cloud \smcb\ (\S\ref{sec:smcb1sed}).
Our principal results are:

\begin{enumerate}
\item The dust IR emission from a 6.25 deg$^2$ region including
      the SMC Bar and Eastern Wing
      is well reproduced by a model with $\dN/d\Urad\propto \Urad^{-1.8}$, 
      $10^{-1.0} \le \Urad \le 10^{2.75}$,
      and $N_{\rm H}^{\rm tot}\approx 5.4\times 10^{21}\cm^{-2}$
      using the SMC dust model of Weingartner \& Draine (2001a).
      This dust has a very low PAH abundance, with C/H $\ltsim$ 0.2ppm
      ($\ltsim 0.4\%$ of the SMC interstellar 
      C abundance) incorporated into PAHs.
      The dust-weighted mean radiation intensity is 
      $\langle U\rangle \approx 2$, but the starlight intensity must
      have a broad distribution ranging from $\Umin\ltsim 0.3$ to
      $\Umax\gtsim 50$ (\S\ref{sec:smcsed}).

\item The IR emission from the molecular cloud \smcb\ can be
      reproduced by a dust model consisting of silicates, graphite, 
      and PAHs with $N_{\rm H}^{\rm tot}\approx 4.2\times 10^{22}\cm^{-2}$
      illuminated by starlight with a power law distribution
      $\dN/d\Urad\propto \Urad^{-2}$, $10^{-0.5} \le \Urad \le 10^{2.5}$  
      (\S\ref{sec:smcb1sed}).
      The required PAH abundance in \smcb\ is 1.5$\ppm$, or 
      $\sim$3\% of the SMC interstellar C abundance.

\item The observed 11.3$\um$/7.7$\um$ and 6.2$\um$/7.7$\um$ band ratios 
      for \smcb\ fall outside the ``allowed'' region 
      predicted for the Milky Way PAH model (see Figures 16, 17 of 
      Draine \& Li 2001).
      In \S\ref{sec:mwmod}, \S\ref{sec:h2c}, 
      and \S\ref{sec:reddenedisrf}, we examine the possibility that 
      this could be due to (1) an enhanced neutral fraction,
      (2) an enhanced H/C ratio for the PAHs in \smcb,
      or (3) reddening of the radiation field to which the PAHs in
      \smcb\ are exposed,
      but none of these scenarios successfully accounts for the
      observed 5--15$\micron$ spectrum. 
      We instead conclude that the intrinsic
      IR band strengths for the PAH mixture in \smcb\ differ from
      the band strengths adopted for Milky Way PAH mixtures
      (\S\ref{sec:ir_band_strengths}).  
      This indicates that the PAH mixture in \smcb\ differs 
      in detail from the PAH mixtures present in a number of
      Milky Way regions.
      The band strengths suggested by the \smcb\ spectrum 
      are close to ``laboratory'' band strengths.

\item The PAH abundance in \smcb\ exceeds that in the
      SMC Bar by a factor $\gtsim 8$ (C/H=1.5 ppm vs $\ltsim$ 0.2ppm).
      While the average SMC extinction curve has no detectable 2175\AA\ hump,
      we predict a 2175\AA\ extinction excess of $\simgt 0.5\magni$ 
      for the dust in \smcb\ (\S\ref{sec:extinction}).

\end{enumerate}

Two types of measurements would be of great value to test our dust model:
(1) UV extinction measurements 
for sightlines through or near \smcb\ to test for the predicted
2175\AA\ feature;
(2) $3-15\micron$ spectrophotometry of other 
regions in the SMC where UV extinction measurements place a strong upper
limit on the 2175\AA\ feature -- the present model predicts that PAH
emission should not be seen from such regions.

\acknowledgments
We thank W.T. Reach for providing us with the ISO spectrum 
of SMC and for helpful comments; D.P. Finkbeiner for help 
in obtaining the DIRBE fluxes for the \smcb\ molecular cloud; 
S. Stanimirovic for obtaining the SMC HI mass; 
G. Helou for valuable comments;
and R.H. Lupton for the availability of 
the SM plotting package. 
This research was supported in part by 
NASA grant NAG5-10811 and NSF grant AST-9988126.

\end{document}